\documentclass[twocolumn,english]{revtex4}
\usepackage[T1]{fontenc}
\usepackage[utf8]{luainputenc}
\usepackage{amsmath}
\usepackage{amssymb}
\usepackage{graphicx}
\usepackage{esint}

\makeatletter
\@ifundefined{textcolor}{}
{%
 \definecolor{BLACK}{gray}{0}
 \definecolor{WHITE}{gray}{1}
 \definecolor{RED}{rgb}{1,0,0}
 \definecolor{GREEN}{rgb}{0,1,0}
 \definecolor{BLUE}{rgb}{0,0,1}
 \definecolor{CYAN}{cmyk}{1,0,0,0}
 \definecolor{MAGENTA}{cmyk}{0,1,0,0}
 \definecolor{YELLOW}{cmyk}{0,0,1,0}
}

\usepackage{babel}

\usepackage{babel}

\usepackage{babel}

\usepackage{babel}

\usepackage{babel}

\usepackage{babel}

\usepackage{babel}

\usepackage{babel}

\usepackage{babel}

\usepackage{babel}

\usepackage{babel}

\usepackage{babel}

\usepackage{babel}

\makeatother

\usepackage{babel}
\begin{document}

\title{Strong eigenstate thermalization hypothesis}

\author{Garry Goldstein and Natan Andrei}

\address{Department of Physics, Rutgers University}

\address{Piscataway, New Jersey 08854}
\begin{abstract}
We present a generalization of the ETH conjecture. Using this generalization
we are able to derive the fact that an arbitrary eigenstate of a general
many body system may be used to represent microcanonical ensemble
in any many body experiment that involves only local operators and
projectors onto eigenstates of the system Hamiltonian. In particular
we extend the ETH to include some non-local operators. We present
a derivation of this conjecture in the case of a many body model whose
Hamiltonian is composed of two parts: an integrable Hamiltonian and
a small but finite Gaussian perturbation. 
\end{abstract}
\maketitle

\section{\label{sec:Introduction}Introduction}

Understanding the non-equilibrium long time dynamics of non-integrable
many body systems has become a major theoretical and experimental
challenge over recent years. This research is driven by the advances
in the field of ultracold atoms \cite{key-1-1,key-3-1,key-4-1,key-5-1,key-6-1,key-7-1,key-8-1,key-9-1,key-11-1}.
These advances allow the detailed experimental study of the time evolution
(of thermally isolated systems) from given initial states under the
influence of well defined Hamiltonians (in particular ones not interacting
with any bath). One of the key questions that is under investigation
is the validity of the Gibbs ensemble formula for describing equilibrium
states of such systems. The ETH hypothesis is one of the cornerstones
in deriving the validity of the Gibbs formula for thermodynamic quantities
and for local operators. It states that for any non-integrable, non
many-body localized Hamiltonian $H$: 
\begin{equation}
Tr\left\{ \left|\lambda\right\rangle \left\langle \lambda\right|O\right\} =F_{O}\left(E_{\lambda}\right)\label{eq:ETH}
\end{equation}
for most eigenstates $\left|\lambda\right\rangle $, $H\left|\lambda\right\rangle =E_{\lambda}\left|\lambda\right\rangle $.
Here $O$ is any local operator and $F_{O}$ is some smooth function
that depends only on the energy of the state $\left|\lambda\right\rangle $,
$E_{\lambda}$, and not on any other of its properties. There currently
does not exist a general proof of the ETH conjecture. There are however
some derivation for special classes of Hamiltonians. In particular
there is a rigorous derivation of the ETH conjecture for integrable
Hamiltonians perturbed by small Gaussian perturbations \cite{key-1}.
Furthermore for nuclear models calculations have shown that individual
wavefunctions reproduce thermodynamic expectation values \cite{key-3}.
There are various proofs that semiclassical quantum systems, whose
classical counterparts are chaotic, satisfy the ETH hypothesis \cite{key-4,key-5,key-6,key-7}.
In particular the low density billiards in the semiclassical limit
satisfy ETH \cite{key-2,key-8}. More generally the ETH conjecture
may be derived from Berry's conjecture which is valid for semiclassical
chaotic systems \cite{key-9}. Furthermore the validity of the ETH
hypothesis has been numerically demonstrated for a variety of non-integrable
models \cite{key-10,key-11,key-12,key-13,key-14,key-15}.

With the ETH hypothesis it is possible to show that the expectation
value with respect to a given state is equal to the one given by the
microcanonical ensemble, e.g. 
\begin{equation}
\left\langle \lambda\right|O\left|\lambda\right\rangle =\frac{1}{N_{E_{\lambda},\delta E}}\sum_{\left|E_{\alpha}-E_{\lambda}\right|<\delta E}\left\langle \alpha\right|O\left|\alpha\right\rangle \label{eq:microcanonical}
\end{equation}
Here $\delta E$ is a small energy width and $N_{E_{\lambda},\delta E}$
is the number of states in the interval $\left[E_{\lambda}-\delta E,\, E_{\lambda}+\delta E\right]$
(this is indeed true since all the entries on the right hand side
of Eq. (\ref{eq:microcanonical}) are the same). Furthermore through
a saddle point argument it is possible to show that for the purposes
of computing local expectation values a single state is equivalent
to the canonical ensemble: 
\begin{equation}
\left\langle \lambda\right|O\left|\lambda\right\rangle =\frac{1}{Z}\sum e^{-\beta E_{\alpha}}\left\langle \alpha\right|O\left|\alpha\right\rangle \label{eq:Canonical}
\end{equation}
where $Z=\sum e^{-\beta E_{\alpha}}$ and $\beta$ is the the inverse
temperature chosen such that $\frac{1}{Z}\sum e^{-\beta E_{\alpha}}E_{\alpha}=E_{\lambda}$.
This is true in the thermodynamic limit since the energies $E_{\alpha}$
that contribute strongly to the right hand side of Eq. (\ref{eq:Canonical})
are strongly peaked about one value $E_{\lambda}$. Using this it
is possible to show that a single eigenstate is equivalent to the
canonical ensemble when the system undergoes an arbitrary evolution
and is subjected to any set of measurements as long as all these operations
are local \cite{key-17}. However it is of great interest to consider
non-local operations. The most important non-local operation is projection
onto an eigenstate of the Hamiltonian $H$: $\Pi_{E}$ which comes
from an accurate measurement of the system energy. To match such measurements
it is important to generalize the ETH formalism to include such projectors.
A sufficient postulate that would extend ETH to include all such energy
projectors is given by: 
\begin{equation}
Tr\left\{ \prod_{i=1}^{n}O_{i}\left|\lambda_{i}\right\rangle \left\langle \lambda_{i}\right|\right\} =F_{O_{1},..O_{n}}\left(Distinct\left\{ E_{\lambda_{1}},...E_{\lambda_{n}}\right\} \right)\label{eq:Strong_ETH}
\end{equation}
The function $F_{O_{1},...O_{n}}$ depends smoothly on its variables
function and the label $Distinct$ means to count only different $\lambda$'s,
e.g. if $\left|\lambda_{i}\right\rangle =\left|\lambda_{j}\right\rangle $
then there is only one energy in the set $Distinct\left\{ E_{\lambda_{1}},...E_{\lambda_{n}}\right\} $
corresponding to the elements $\lambda_{i}$ and $\lambda_{j}$. We
note that the case $n=1$ corresponds to the usual ETH hypothesis
see Eq. (\ref{eq:ETH}). Using Eq. (\ref{eq:Strong_ETH}) it is possible
to prove the ETH hypothesis, Eq. (\ref{eq:ETH}), in the case when
$O$ contains non-local parts in the form of projectors onto eigenstates
of the Hamiltonian $H$. The rest of the paper is organized as follows.
In Section \ref{sec:Disordered-Hermitian-matricies} we show the strong
ETH result is valid in the case of a integrable model with a Gaussian
perturbation. In Section \ref{sec:Applications} we show that for
systems that satisfy Eq. (\ref{eq:Strong_ETH}) a single eigenstate
is equivalent to the microcanonical ensemble for any possible experiment
that uses only local operations and projectors (we also prove Eq.
(\ref{eq:ETH}) when $O$ includes projectors). To show equivalence
to the canonical ensemble we need some more assumptions. Also as an
application in Section \ref{sub:Linear-Response} we use Eq. (\ref{eq:Strong_ETH})
show that the linear response function of a single eigenstate is identical
to the linear response of the canonical ensemble. In Section \ref{sec:Conclusions}
we conclude.

\section{\label{sec:Disordered-Hermitian-matricies}Disordered Hermitian matrices}

We would like to derive Eq. (\ref{eq:Strong_ETH}) for a simple model.
For simplicity we will prove it in the case when $\left|\lambda_{1}\right\rangle \neq\left|\lambda_{2}\right\rangle \neq....\neq\left|\lambda_{n}\right\rangle $
(the case when some of the eigenvectors are identical is handled highly
similarly). We will consider a model where the Hamiltonian is composed
of two part, an integrable part and a Gaussian perturbation, e.g.
\begin{equation}
H=\sum_{i}E_{i}\left|i\right\rangle \left\langle i\right|+H_{Gauss}\label{eq:Hamiltonian}
\end{equation}

Here it is assumed that the $E_{i}$ are dense with average level
spacing $\Delta$ and $H_{Gauss}$ has matrix elements $h_{ij}\left|i\right\rangle \left\langle j\right|$
that have gaussian distribution with $\left\langle h_{ij}h_{kl}\right\rangle =\varepsilon^{2}\delta_{il}\delta_{jk}$.
Let $c_{ij}$ be the unitary matrix that diagonalizes $H$, e.g. $c^{\dagger}Hc=D$
for a diagonal matrix $D$. The the expectation value of the expression
in Eq. (\ref{eq:Strong_ETH}) is given by: 
\begin{equation}
\begin{array}[t]{l}
Tr\left\{ \prod_{\eta=1}^{n}O_{\eta}\left|\lambda_{\eta}\right\rangle \left\langle \lambda_{\eta}\right|\right\} =\\
=\sum_{i_{1},..i_{n},j_{i},..j_{n}}Cyc\left\{ \prod_{\eta=1}^{n}\left\langle j_{\eta+1}\right|O_{\eta}\left|i_{\eta}\right\rangle c_{i_{\eta}\lambda_{\eta}}c_{\lambda_{\eta+1}j_{\eta+1}}^{*}\right\} 
\end{array}\label{eq:Operator_Value}
\end{equation}
where $Cyc$ means to cyclically identify $n+1=1$. Now the probability
distribution of the expression given in Eq. (\ref{eq:Operator_Value})
for eigenenergies $E_{\lambda_{1}},...E_{\lambda_{n}}$ is given by:
\begin{equation}
\begin{array}[t]{l}
P\left\{ Tr\left\{ \prod_{\eta=1}^{n}O_{\eta}\left|\lambda_{\eta}\right\rangle \left\langle \lambda_{\eta}\right|\right\} \right\} \\
\propto\sum_{I_{i},..I_{n}}\int dc_{ij}\delta\left(c^{\dagger}c-I\right)\prod_{i<j}\delta\left(\left(c^{\dagger}Hc\right)_{ij}\right)\times\\
\times\prod_{I_{1},..I_{n}}\delta\left(\left(c^{\dagger}Hc\right)_{I_{i}I_{i}}-E_{\lambda_{i}}\right)\times\\
\times\prod\exp\left(-\frac{1}{\varepsilon^{2}}\sum_{i<j}\left|H_{ij}\right|^{2}-\frac{1}{2\varepsilon^{2}}\sum_{i}\left(H_{ii}-E_{i}\right)^{2}\right)\times\\
\sum_{i_{1},..i_{n},j_{i},..j_{n}}Cyc\left\{ \prod_{\eta=1}^{n}\left\langle j_{\eta+1}\right|O_{\eta}\left|i_{\eta}\right\rangle c_{i_{\eta}\lambda_{\eta}}c_{\lambda_{\eta+1}j_{\eta+1}}^{*}\right\} 
\end{array}\label{eq:Probability distribution}
\end{equation}

The form of expression in the exponential (which is extensive in the
system size): $\prod\exp\left(-\frac{1}{\varepsilon^{2}}\sum_{i<j}\left|H_{ij}\right|^{2}-\frac{1}{2\varepsilon^{2}}\sum_{i}\left(H_{ii}-E_{i}\right)^{2}\right)$
gives the probability distribution in Eq. (\ref{eq:Probability distribution})
a very small spread, with the distribution having width $\sim\sqrt{\frac{\Delta}{\varepsilon}}$.
Therefore for fixed energies $E_{\lambda_{1}},...E_{\lambda_{n}}$
the expression in Eq. (\ref{eq:Probability distribution}) has uncertainty
that scales like $\sim\sqrt{\frac{\Delta}{\varepsilon}}$ and tends
to zero exponentially in the thermodynamic limit. The expression in
Eq. (\ref{eq:Strong_ETH}) depends therefore only on the energies
$E_{\lambda_{1}},...E_{\lambda_{n}}$ and Eq. (\ref{eq:Strong_ETH})
is proven.

We would like to note that using a very similar procedure it is possible
to prove that for any local operators $O_{1},...O_{n}$ and generic
Hamiltonians $H_{1},...H_{n}$ chosen from the distribution given
in Eq. (\ref{eq:Hamiltonian}) we have that: 
\begin{equation}
Tr\left\{ \prod_{i=1}^{n}O_{i}\left|\lambda_{i}\right\rangle \left\langle \lambda_{i}\right|\right\} =F_{O_{1},..O_{n}}\left(\left\{ E_{\lambda_{1}},...E_{\lambda_{n}}\right\} \right)\label{eq:Strong_ETH-1}
\end{equation}

for most states such that $H_{i}\left|\lambda_{i}\right\rangle =E_{\lambda_{i}}\left|\lambda_{i}\right\rangle $.

\section{\label{sec:Applications}Applications}

We would like to show that for any system that satisfies Eq. (\ref{eq:Strong_ETH})
a single eigenstate is completely equivalent to the microcanonical
ensemble for the purposes of doing any experiment that involves only
local Hamiltonians and projectors $\Pi_{E}$ onto eigenstates of the
Hamiltonian $H$. As a warm up we will show that the linear response,
e.g. conductivity of an eigenstate is equivalent to that of the canonical
ensemble, even though this measurement involves projectors.

\begin{figure}
\begin{centering}
\includegraphics[width=1\columnwidth]{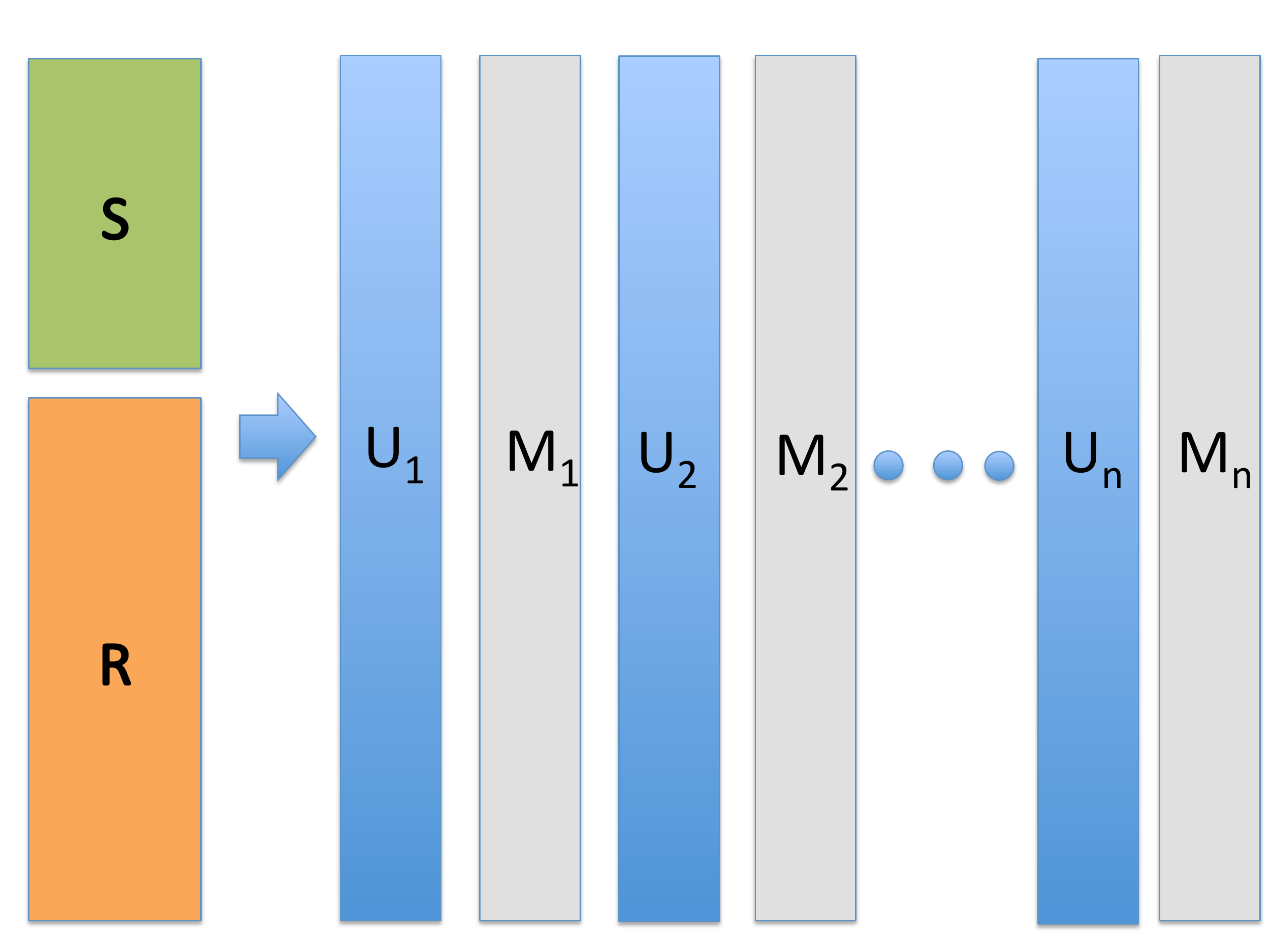} 
\par\end{centering}

\protect\protect\protect\protect\protect\protect\caption{\label{fig:Circuit} A generic experiment. The system ($S$) is entangled
with a reservoir ($R$) and undergoes multiple rounds of evolution
$U_{i}$ and measurement $M_{i}$. }
\end{figure}

\subsection{\label{sub:Linear-Response}Linear Response}

We would like to show that the linear response of an eigenstate in
a model satisfying Eq. (\ref{eq:Strong_ETH}) has the same linear
response as the canonical and microcanonical ensemble. We note that
for many operators it is possible to do this directly from the ETH,
see Eq. (\ref{eq:ETH}). To do so recall that the linear response
function $D_{O_{1},O_{2}}^{\omega}\left(\lambda\right)$ of a system
initialized in an eigenstate $H\left|\lambda\right\rangle =E_{\lambda}\left|\lambda\right\rangle $
at $t=0$ has the following definition. Suppose the system is subjected
to the time dependent Hamiltonian $\tilde{H}\left(t\right)=H+ae^{-i\omega t}O_{1}$
for some infinitesimal constant $a$ and an arbitrary operator $O_{1}$,
then it develops a non-zero expectation value for the operator $O_{2}$
given by $\left\langle O_{2}\left(t\right)\right\rangle =aD_{O_{1},O_{2}}^{\omega}\left(\lambda\right)e^{-i\omega t}$
at long times. The linear response coefficient may be calculated,
it is given by \cite{key-16}: 
\begin{align}
D_{O_{1},O_{2}}^{\omega}\left(\lambda\right) & =\sum_{\chi}\frac{\left\langle \lambda\right|O_{2}\left|\chi\right\rangle \left\langle \chi\right|O_{1}\left|\lambda\right\rangle }{\omega-\left(E_{\chi}-E_{\lambda}\right)+i\epsilon}+\nonumber \\
 & +\sum_{\chi}\frac{\left\langle \lambda\right|O_{1}\left|\chi\right\rangle \left\langle \chi\right|O_{2}\left|\lambda\right\rangle }{-\omega-\left(E_{\chi}-E_{\lambda}\right)-i\epsilon}\label{eq:Linear_response}
\end{align}
Here $\epsilon$ is an infinitesimal positive number and $\left|\chi\right\rangle $
is a complete set of states for the Hamiltonian $H$. From the form
of Eq. (\ref{eq:Linear_response}) we see that based on Eq. (\ref{eq:Strong_ETH})
if two states $\left|\lambda_{1}\right\rangle $ and $\left|\lambda_{2}\right\rangle $
have the same energy then the linear response coefficients $D_{O_{1},O_{2}}^{\omega}\left(\lambda_{1}\right)=D_{O_{1},O_{2}}^{\omega}\left(\lambda_{2}\right)$.
From this it is straightforward to obtain that: 
\begin{equation}
D_{O_{1},O_{2}}^{\omega}\left(\lambda\right)=D_{O_{1},O_{2}}^{\omega}\left(\rho_{micro}\right)=D_{O_{1},O_{2}}^{\omega}\left(\rho_{can}\right)\label{eq:Linear_response_canonical}
\end{equation}
namely, a single eigenstate has the same linear response as the canonical
and microcanonical ensembles. We note that in the case when $O_{2}=O_{1}^{\dagger}$
the imaginary part of the expression in Eq. (\ref{eq:Linear_response})
is given by: 
\begin{align}
Im\left\{ D_{O,O^{\dagger}}^{\omega}\left(\lambda\right)\right\}  & =-\pi\sum_{\chi}\left|\left\langle \chi\right|O\left|\lambda\right\rangle ^{2}\right|\delta\left(\omega-\left(E_{\chi}-E_{\lambda}\right)\right)\nonumber \\
 & +\pi\sum_{\chi}\left|\left\langle \chi\right|O\left|\lambda\right\rangle ^{2}\right|\delta\left(\omega+\left(E_{\chi}-E_{\lambda}\right)\right)\label{eq:Absobtion}
\end{align}
This is a highly non-local term that involves a projector $\Pi_{E_{\lambda}\pm\omega}$
onto eigenstates of the Hamiltonian $H$, exactly the scenario our
formalism handles. We note that unlike the usual ETH, with Eq. (\ref{eq:Strong_ETH})
this result is valid even when the local operators $O_{1},O_{2}$
are far apart in real space so that their product is no longer a local
operator.

\subsection{\label{sub:General-experiment}General experiment}

We would like to show that a single eigenstate will produce the same
measurement outcomes as the microcanonical ensemble for any experiment
that involves only local evolution and measurement as well as any
projectors $\Pi_{E}$ onto eigenstates of the Hamiltonian $H$. Indeed
the most generic experiment \cite{key-17} may be described as preparing
the system $S$ in some state and preparing an auxiliary reservoir
$R$ in a well defined initial state $\rho_{R}$, then having the
system, reservoir ensemble undergo rounds of evolution $U_{i}$ and
measurement $M_{i}$ with the measurement operators being given by
$M_{i,j}$ \cite{key-17}. The probability of some outcome is given
by: 
\begin{equation}
Tr\left\{ M_{n,i_{n}}U_{n}....M_{1,i_{1}}U_{1}\left|\lambda\right\rangle \left\langle \lambda\right|\otimes\rho_{R}U_{1}^{\dagger}M_{1,i_{1}}^{\dagger}....U_{n}^{\dagger}M_{n,i_{n}}^{\dagger}\right\} \label{eq:Measurement}
\end{equation}

By inserting multiple resolutions of identity $\sum_{\chi}\left|\chi\right\rangle \left\langle \chi\right|$
between all the operators in Eq. (\ref{eq:Measurement}) it is possible
to show that the probability of the outcome $M_{1,i_{1}}....M_{n,i_{n}}$
depends only on the energy of the eigenstate $\left|\lambda\right\rangle $:
$E_{\lambda}$. This proves that the microcanonical ensemble is equivalent
to a single eigenstate for any experiment involving only local operations
and projectors $\Pi_{E}$. In a very similar manner we can prove Eq.
(\ref{eq:ETH}) for $O$ composed of any local operators and projectors
onto the eigenstates of $H$. We note also that unlike the regular
ETH our formalism allows for the measurement operators $M_{k,i_{k}}$
to be far apart in real space so that their product is non-local.
We note that to prove equivalence of a single state to the canonical
ensemble, which has some energy spread, we need to further assume
that the projectors $\Pi_{E_{i}}$ and $\Pi_{E_{j}}$ come with an
integration over energy so that the total operator depends only on
projectors in the combination $E_{i}-E_{j}$ and any local operators.
We would like to note that similar results are possible for experiments
that involve only local operators and projectors onto eigenstates
of generic Hamiltonians that satisfy Eq. (\ref{eq:Strong_ETH-1})
above.

\section{\label{sec:Conclusions}Conclusions}

In this work we have considered extensions of the ETH hypothesis to
include some nonlocal operators, namely projectors onto eigenstates
of the Hamiltonian of the system. We extended the ETH hypothesis to
include multiple eigenstates thereby proving the usual ETH including
the case of projectors and measurements. We supported our results
by considering a simple model of an integrable Hamiltonian and a Gaussian
perturbation. It is of great interest to extend the ETH to other non-local
operators, it is a direction of research the authors are currently
pursuing.

\textbf{Acknowledgments}: This research was supported by NSF grant
DMR 1410583 and Rutgers CMT fellowship.

\end{document}